# Spatially indirect intervalley excitons in bilayer WSe$_2$

Zhiheng Huang[1,2,†], Yanchong Zhao[1,2,†], Tao Bo[1,3,†], Yanbang Chu[1,2], Jinpeng Tian[1,2], Le Liu[1,2], Yalong Yuan[1,2], Fanfan Wu[1,2], Jiaojiao Zhao[1,2], Lede Xian[3], Kenji Watanabe[4], Takashi Taniguchi[5], Rong Yang[1,3,6], Dongxia Shi[1,2,6], Luojun Du[7]*, Zhipei Sun[7,8], Sheng Meng[1,2,3]*, Wei Yang[1,2,6]*, Guangyu Zhang[1,2,3,6]*

[1]Beijing National Laboratory for Condensed Matter Physics; Key Laboratory for Nanoscale Physics and Devices, Institute of Physics, Chinese Academy of Sciences, Beijing, 100190, China

[2]School of Physical Sciences, University of Chinese Academy of Sciences, Beijing 100190, China

[3]Songshan Lake Materials Laboratory, Dongguan, Guangdong Province 523808, China

[4]Research Center for Functional Materials, National Institute for Materials Science, 1-1 Namiki, Tsukuba 305-0044, Japan

[5]International Center for Materials Nanoarchitectonics, National Institute for Materials Science, 1-1 Namiki, Tsukuba 305-0044, Japan

[6]Beijing Key Laboratory for Nanomaterials and Nanodevices, Beijing 100190, China

[7]Department of Electronics and Nanoengineering, Aalto University, Tietotie 3, FI-02150, Finland

[8]QTF Centre of Excellence, Department of Applied Physics, Aalto University, FI-00076 Aalto, Finland

†These authors contributed equally to this work: Zhiheng Huang, Yanchong Zhao, Tao Bo

*Corresponding authors. Email: luojun.du@aalto.fi; smeng@iphy.ac.cn; wei.yang@iphy.ac.cn; gyzhang@iphy.ac.cn

**Spatially indirect excitons with displaced wavefunctions of electrons and holes play a pivotal role in a large portfolio of fascinating physical phenomena and emerging optoelectronic applications, such as valleytronics, exciton spin Hall effect, excitonic integrated circuit and high-temperature superfluidity. Here, we uncover three types of spatially indirect excitons (including their phonon replicas) and their quantum-confined Stark effects in hexagonal boron nitride encapsulated bilayer $WSe_2$, by performing electric field-tunable photoluminescence measurements. Because of different out-of-plane electric dipole moments, the energy order between the three types of spatially indirect excitons can be switched by a vertical electric field. Remarkably, we demonstrate, assisted by first-principles calculations, that the observed spatially indirect excitons in bilayer $WSe_2$ are also momentum-indirect, involving electrons and holes from Q and K/Γ valleys in the Brillouin zone, respectively. This is in contrast to the previously reported spatially indirect excitons with electrons and holes localized in the same valley. Furthermore, we find that the spatially indirect intervalley excitons in bilayer $WSe_2$ can exhibit considerable, doping-sensitive circular polarization. The spatially indirect excitons with momentum-dark nature and highly tunable circular polarization open new avenues for exotic valley physics and technological innovations in photonics and optoelectronics.**

Exciton, a hydrogen-atom-like electron-hole pair bound by their mutual Coulomb interaction, plays an important role in a wide variety of intriguing optoelectronic properties of materials [1-4]. Depending on whether the wavefunctions of electrons and holes are spatially separated, excitons can be divided into two types: spatially direct and indirect excitons. Because of the separation of the electrons and holes, spatially indirect excitons have a much longer lifetime than spatially direct excitons and are predicted to exhibit a wide spectrum of emergent physical phenomena, including but not limit to quantum-confined Stark effect [5-7], Bose-Einstein condensation [8-12], strongly correlated excitonic insulator states [13-15], high-temperature superconductivity [16], valley physics [17-19] and dissipationless exciton transistors [20-22]. The recent emergence of two-dimensional transition metal dichalcogenides (TMDCs) and their van der Waals (vdW) heterostructures offers an unprecedented platform to realize spatially indirect excitons. Indeed, spatially indirect excitons have thus far been demonstrated in a wide variety of TMDC homo- and hetero-structures [17], such as $MoS_2/WS_2$ [23-25], $MoS_2/WSe_2$ [26,27], $MoSe_2/WSe_2$ [18,20,28-31] and bilayer $MoS_2$ [7,32-37]. Specially, owing to the strongly reduced dielectric screening, spatially indirect excitons in home-/hetero-bilayers of TMDCs possess substantial binding energies and show crucial advantages for applications, for example, superfluidity at high-temperature [9].

To date, the studies of spatially indirect excitons have mainly focused on the momentum-bright species with electrons and holes localized in the same valley of the Brillouin zone (BZ) [7,17-23,27-40]. On the other hand, because of the existence of multiple electronic valleys, TMDC homo- and hetero-bilayers can also exhibit spatially indirect excitons with momentum-dark nature (that is, electrons and holes are from different valleys of the BZ) [6,26,41-43]. Since electrons and holes are further separated in momentum-space, spatially indirect intervalley excitons, in principle, can possess a longer lifetime than spatially indirect but momentum-direct excitons, and represent an advantageous scenario for numerous theoretical, experimental and technological advances. However, in contrast to the well-studied spatially indirect excitons with momentum-bright feature, experimental progress on spatially indirect intervalley excitons is still largely limited.

In this work, we demonstrate three types of spatially indirect intervalley excitons (i.e., two Q-K transitions, one Q-Γ exciton and their phonon replicas) and their quantum-confined Stark effects in hexagonal boron nitride (*h*-BN) encapsulated bilayer WSe$_2$, through the combination of electric field-dependent photoluminescence (PL) measurements and density functional theory (DFT) calculations. The energy order between the three types of spatially indirect intervalley excitons can be switched by an electric field, owing to their different electric dipole moments. Interestingly, these spatially indirect intervalley excitons in bilayer WSe$_2$ show considerable negative circular polarization that is highly tunable with electron doping. Our results not only provide a complete understanding of the puzzling multiplet emissions in WSe$_2$ bilayers, but also present the new possibilities for valleytronics, high-temperature superfluidity and advanced functionalities in photonics and optoelectronics.

Among various TMDCs, bilayer WSe$_2$ provides a promising platform for spatially indirect intervalley excitons. First, for bilayer WSe$_2$, the conduction band minimum is located at the Q (Q′) points of the BZ, while the critical points of valence band are at K/K′ and Γ, as shown in Fig. 1(a) [44-47]. Consequently, the lowest exciton transition in bilayer WSe$_2$ should be momentum-indirect Q-K or Q-Γ excitons, in marked contrast to the momentum-direct K-K transition in monolayer case. Second, as the Bloch states at conduction band Q, valence band K and Γ have distinct orbital compositions [Fig. 1(a)], their wavefunctions show different interlayer hybridization and reside at different positions in real-space [Fig. 1(b)] (Supplemental Materials [48]) [45,49]. Therefore, momentum-indirect Q-K and Q-Γ excitons are also spatially indirect with finite out-of-plane electric dipole moments. Third, because of the substantial exciton-phonon coupling and the inevitable existence of defects [50-55], Q-K and Q-Γ transitions in bilayer WSe$_2$, in principle, can be activated by phonon/defect scattering and show strong PL responses. Although there have been some studies on spatially indirect intervalley excitons in bilayer WSe$_2$ [6,41,56-59], their underlying origin remains equivocal. In addition, previous research has reported only one type of Q-K exciton [6,41], the other type of Q-K transition and the Q-Γ exciton have not been revealed.

We fabricate high quality *h*-BN encapsulated bilayer WSe$_2$ devices by a vdW mediated dry transfer method (see Supplemental Materials [48] for more details). Few-layer graphene (FLG) is used as both the bottom and top gate electrodes to further screen the charged impurities on SiO$_2$ substrates and improve the device quality [Fig. 1(c)]. Three *h*-BN encapsulated bilayer WSe$_2$ devices (labelled as D1, D2 and D3) are studied, showing similar behavior (see Supplemental Materials [48] for more details). Unless otherwise specified, the data presented here are taken from device D1 in a high vacuum at 10 K, excited by 1.96 eV (633 nm) radiation. The dual-gated devices enable us to independently tune the vertical electric field ($E_z$) and doping density ($n_0$) (Supplemental Materials [48]).

Figure 2(a) shows the PL spectrum of bilayer WSe$_2$ without applying gate voltages. Apart from the momentum-direct K-K transitions at around 1.69 eV ($X_0$), seven lower energy peaks in the range of 1.50 eV to 1.65 eV (black dotted box), corresponding to the momentum-indirect transitions, can be clearly observed [6,57]. It is noteworthy that benefiting from the high quality of our samples, the number of momentum-indirect excitons revealed here is larger than that previously observed [6]. As we mentioned above, the momentum-indirect excitons in bilayer WSe2 should also be spatially indirect. To confirm this, we perform electric field-tunable PL measurements. Figure 2(b) depicts the color plot of PL spectra as a function of $E_z$. Obviously, all the momentum-indirect excitons are highly tunable with $E_z$, evidencing the quantum-confined Stark effects and their spatially indirect nature. Note that the emission energy of K-K transition $X_0$ remains unchanged with $E_z$ (Supplemental Materials [48]). To better resolve the fine features, we plot the first-order derivative of intensity ($dI/dE$) [Fig. 2(c)]. Figure 2(d) displays the energies of different spatially indirect intervalley emissions as a function of $E_z$, extracted from Fig. 2(c). The spatially indirect intervalley excitons in bilayer WSe2, at first glance, can be divided into two types: one [blue dashed lines in Fig. 2(d)] with cross-shape features and the other [red dashed lines in Fig. 2(d)] with a conversion from nonlinear Stark shift at small $|E_z|$ to linear Stark shift at large $|E_z|$.

We tentatively assign the former (latter) type of spatially indirect intervalley excitons as Q-Γ (Q-K) transitions. Note that here we use Q-K transitions to denote all the possible transitions between electrons at Q/Q′ and holes at K/K′ and the same for Q-Γ transitions. We extract the vertical displacement of these excitons from Fig. 2(c) using $d_\perp = -\frac{dE}{e \cdot dE_z}$, where $E$ is the emission energy and $e$ is the elementary charge. Note that the sign of $d_\perp$ represents the direction of electric dipole moment: positive (negative) means vertical upward (downward). For Q-Γ transitions, the $d_\perp$ is nearly fixed at ±1.40 Å [purple dots in Fig. 3(a)]. For Q-K transitions, the $d_\perp$ is about ±1.80 Å at zero electric field, then it gradually increases with the electric field, and finally it saturates at ±4.50 Å [yellow dots in Fig. 3(b)].

To support our assignment, we then perform DFT calculations to derive the equivalent positions of spin-up/-down wavefunctions at conduction band Q, valence band K and Γ. The equivalent position of a wavefunction is defined as $r_z = \int_{-\infty}^{+\infty} r|\varphi(r)|^2 dr$, where $|\varphi(r)|^2$ denotes the probability density of wavefunction $\varphi(r)$ at position $r$. The origin point (positive direction) is set as the midpoint between the two layers (vertical upward). For spin-up wavefunctions at conduction band Q (Q′), valence band K (K′) and Γ, the calculated equivalent positions at zero electric field are $r_z = -0.22t$ ($0.22t$), $-0.48t$ ($0.48t$) and 0, respectively [Fig. 1(b)], where $t = 6.6$ Å is the interlayer distance of bilayer WSe2. For spin-down wavefunctions, the equivalent positions can be obtained simply by time-reversal symmetry (Supplemental Materials [48]). It is worth noting that the equivalent positions of $\pm 0.48t$ indicate the virtually suppressed interlayer hybridization and spin-layer locking for holes at K/K′ [60-62].

For Q-Γ transitions, there are two paths with equal transition probability [Fig. 3(c)]. One is from spin-up electrons at Q to holes at Γ with $d_\perp = r_z(\Gamma) - r_z(Q_\uparrow) = 1.45$ Å and another is from spin-down electrons at Q to holes at Γ with $d_\perp = r_z(\Gamma) - r_z(Q_\downarrow) = -1.45$ Å. Here we take transitions from Q to Γ as an example, transitions from Q′ to Γ could give the same results (Supplemental Materials [48]). Remarkably, $d_\perp$ obtained by first-principles calculations (±1.45 Å) is in good agreement with the experiments (±1.40 Å), confirming our assignment of Q-Γ excitons [Fig. 3(a)].

For Q-K transitions (here we focus on spin-up holes at K valley), there are four possible transition paths [Fig. 3(d)], depending on the spin and valley configuration of carriers. It is noteworthy that spatially indirect intervalley excitons with spin-triplet configuration in bilayer WSe$_2$ may be bright because of the broken out-of-plane mirror symmetry [63]. Among the four possible transitions, two of them (i.e., transitions associated with spin-down electrons at Q and spin-up electrons at Q′) have a large $d_\perp$ ($\sim -0.70t = -4.62$ Å), while the other two (i.e., transitions associated with spin-up electrons at Q and spin-down electrons at Q′) have a small $d_\perp$ ($\sim -0.28t = -1.85$ Å) [Fig. 3(d)]. According to the spatial inversion symmetry, we can know that, for spin-up holes at K′ valley, there are also four possible transitions, but with opposite $d_\perp$: two of them with a large positive $d_\perp$ (~4.62 Å) and the other two with a small positive $d_\perp$ (~1.85 Å) (Supplemental Materials [48]). For Q-K(K′) transitions associated with spin-down holes, we can obtain similar results (Supplemental Materials [48]). Again, a perfect agreement between theoretically calculated values and experimental results is obtained: $d_\perp = \pm 1.85$ Å and $d_\perp = \pm 4.62$ Å obtained by DFT calculations match well the experiments under zero electric field (±1.80 Å) and large electric fields (±4.50 Å) [Fig. 3(b)]. Note that first-principles calculations show that $d_\perp$ change slightly with $E_z$ [Figs. 3(a) and 3(b)]. For example, the large $d_\perp$ of Q-K transition changes from ±4.62 Å at zero electric field to ±4.50 Å at $E_z = 0.2\ V/nm$, which is more consistent with our experimental results [Fig. 3(b)]. In short, we reveal three types of spatially indirect intervalley excitons: two Q-K transitions with different $d_\perp$ and one Q-Γ exciton, providing a complete understanding of the multiplet emissions in bilayer WSe$_2$.

Remarkably, our results manifest three unique features for these spatially indirect intervalley excitons. First, from the comparison of experimental results and first-principles calculations, it can be known that Q-K transition is dominated by the one with small $d_\perp$ at $E_z = 0$, and then gradually becomes dominated by the one with large $d_\perp$ as $|E_z|$ increases. Such exotic characteristic of Q-K transition can be understood as follows. When $E_z = 0$, Q-K transitions with small and large $d_\perp$ share the same energy. However, since the wavefunctions of electrons and holes overlap more, Q-K transition with small $d_\perp$ would acquire larger transition probability and thus dominate the emission. When $E_z$ is applied, the larger Stark shift would lead the Q-K transition with large $d_\perp$ to having a lower energy and thus more occupancy than that with small $d_\perp$. Consequently, Q-K transition with large $d_\perp$ would gain an increasing contribution and eventually dominate the emission under a strong electric field (e.g., $|E_z| > 0.1V/nm$). Note that under an intermediate electric field, the emission is a mixed state, contributed by both Q-K transitions with large and small $d_\perp$. Second, for both Q-K and Q-Γ transitions, there are a series of replicas, labeled as $X_{QK}$ ($X_{Q\Gamma}$), $X_{QK}^1$ ($X_{Q\Gamma}^1$) and $X_{QK}^2$ ($X_{Q\Gamma}^2$ and $X_{Q\Gamma}^3$) in sequence of decreasing emission energy. For the two sets of highest-energy transitions (i.e., $X_{QK}$ and $X_{Q\Gamma}$), the emission intensities are much darker than that of their replicas at lower energy (i.e., $X_{QK}^{1,2}$ and $X_{Q\Gamma}^{1,2,3}$) [Fig. 2(b)], indicating that $X_{QK}/X_{Q\Gamma}$ and $X_{QK}^{1,2}/X_{Q\Gamma}^{1,2,3}$ have different origins. We tentatively attribute $X_{QK}$ ($X_{Q\Gamma}$) and $X_{QK}^{1,2}$ ($X_{Q\Gamma}^{1,2,3}$) to primary Q-K (Q-Γ) transitions activated by defect scattering and their phonon replicas, respectively. Notably, the energy difference (~42 meV) between the primary $X_{QK}$ ($X_{Q\Gamma}$) and the phonon replica $X_{QK}^2$ ($X_{Q\Gamma}^3$) that dominates the emission outstrips the single phonon energy in WSe$_2$ (~37 meV) [54,64]. This indicates that phonon replicas come mainly from two-/multi-phonon scattering, rather than one-phonon scattering. One plausible reason is that two-/multi-phonon processes possess more scattering paths than one-phonon scattering. Third, $X_{Q\Gamma}$ is ~18 meV lower than $X_{QK}$, under zero electric field. This seems a counter-intuitive result because the valence Γ valley is located below the valence K valley [Fig. 1(a)] [6,47], which makes it natural to expect $X_{Q\Gamma}$ to have higher emission energy than $X_{QK}$. In fact, the observed transition energy is determined by the difference between the electronic band-gap and the exciton binding energy, rather than electronic band-gap only. Since the effective mass of holes at Γ point (~1.01 $m_e$; $m_e$ is the free electron mass) is much larger than that at K point (~0.27 $m_e$) [44,58], $X_{Q\Gamma}$ possesses a larger binding energy than $X_{QK}$ and thus it can become the lower energy excitonic state.

Finally, we study the valley properties of spatially indirect intervalley excitons in bilayer WSe$_2$. Figure 4(a) shows the helicity-resolved PL spectra of device D2 for co-circularly (red) and cross-circularly polarized detections (black), excited by $\sigma^+$ radiation. We quantify the degree of circular polarization as $\text{DOP} = \frac{I_{co}-I_{cross}}{I_{co}+I_{cross}}$, where $I_{co}$ and $I_{cross}$ denote the intensities detected under co- and cross-circularly polarized configurations, respectively. Figure 4(b) shows the DOP against the photon energy: the blue line is calculated directly from the measured intensities, while the orange dots present the DOP of spatially indirect intervalley excitons calculated from the fitting intensities. It is explicit that DOPs calculated from the measured and fitted intensities agree well with each other. Thus, for simplicity, all the following DOPs are calculated directly with the measured intensities. Obviously, both Q-K and Q-Γ transitions evince considerable negative DOP (~ − 0.2), whereas the momentum-direct K-K transition has a positive DOP, indicating the opposite chirality between them. Furthermore, we find that the DOP of spatially indirect intervalley excitons in bilayer WSe$_2$ is highly tunable with doping density $n_0$ [Fig. 4(c)]. Figure 4(d) shows the DOP as a function of $n_0$, calculated with the integrated intensity from 1.45 eV to 1.60 eV. The DOP almost keeps constant for hole doping, but gradually vanishes with increasing electron doping density. Such negative, highly tunable circular polarization of spatially indirect intervalley excitons may provide novel device paradigms to exploit the valley degree of freedom other than K (e.g., Q and Γ). In-depth theoretical studies, however, are required to fully understand the optical selection rules and further the highly tunable negative circular polarization of spatially indirect intervalley excitons in bilayer WSe$_2$.

In summary, we reveal three types of spatially indirect intervalley excitons (i.e., two Q-K transitions, one Q-Γ exciton and their phonon replicas) and their giant Stark shift in bilayer WSe$_2$ encapsulated by $h$-BN. Owing to their different electric dipole moments, the energy order and dominant luminescence between the three types of spatially indirect intervalley excitons can be switched by a vertical electric field. Remarkably, these spatially indirect intervalley excitons in bilayer WSe$_2$ show considerable negative circular polarization that is highly tunable with doping density. Our results not only provide a deep understanding of the multiplet momentum-dark emissions in bilayer WSe$_2$, but also hold a promising future for dissipationless exciton transport, high-temperature superfluidity and valley-functional optoelectronic devices with multiple quantum degrees of freedom.

During the preparation of the manuscript, we became aware of a similar work by Mashael M. Altaiary et al. [65].

**Acknowledgments**

We thank Yang Xu and Ting Wang in IOP for useful discussion. This research was supported by the National Key Research and Development Program (Grant Nos. 2020YFA0309600 and 2018YFA0306900), the NSFC (Grants Nos. 61888102, 11834017, and 12074413), the Strategic Priority Research Program of CAS (grant Nos. XDB30000000 and XDB33000000), the Key-Area Research and Development Program of Guangdong Province (Grant No. 2020B0101340001). L.D. gratefully acknowledges the financial support by Academy of Finland (Grant No. 3333099). Z.S. acknowledges support from Academy of Finland (Grant Nos. 314810, 333982, 336144, and 336818), Academy of Finland Flagship Programme (Grant No. 320167, PREIN), the European Union's Horizon 2020 research and innovation program (Grant Nos. 820423, S2QUIP; 965124, FEMTOCHIP), the EU H2020-MSCARISE-872049 (IPN-Bio), and ERC (Grant No. 834742). K.W. and T.T. acknowledge support from the Elemental Strategy Initiative conducted by the MEXT, Japan, Grant Number JPMXP0112101001, JSPS KAKENHI Grant Number JP20H00354 and the CREST(JPMJCR15F3), JST.

**Author contributions**

G.Z. and W.Y. supervised this work; Z.H. and Y.Z. conceived the project and designed the experiments; Z.H. fabricated the devices and carried out the optical measurements; T.B. and S.M. conducted the first-principles calculations; K.W. and T.T. contributed high quality *h*-BN crystals; Z.H., Y.Z., L.D. and W.Y. analyzed the data; Z.H., Y.Z., L.D., W.Y. and G.Z. co-wrote the manuscript. All authors discussed the results and commented on the paper.

**Competing interests**

The authors declare that they have no competing interests.

**Data and materials availability**

All data needed to evaluate the conclusions in the paper are present in the paper and/or the Supplemental Materials. Additional data related to this paper may be requested from the authors.

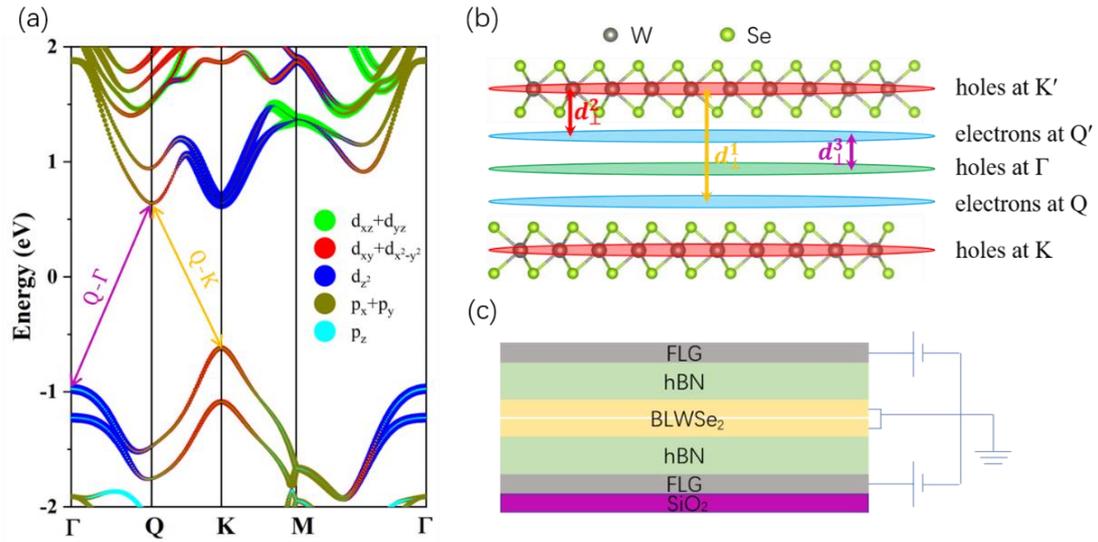

FIG. 1. (a) Orbital projected band structures of bilayer WSe$_2$. Symbol size is proportional to its population in corresponding state. (b) Schematic image of the equivalent positions of spin-up wavefunctions at K/K′ (red ellipses), Q/Q′ (blue ellipses) and Γ points (green ellipse) in real space. $d_\perp^{1-3}$ denote vertical distances between different wavefunctions. (c) Schematic of $h$-BN encapsulated bilayer WSe$_2$ device.

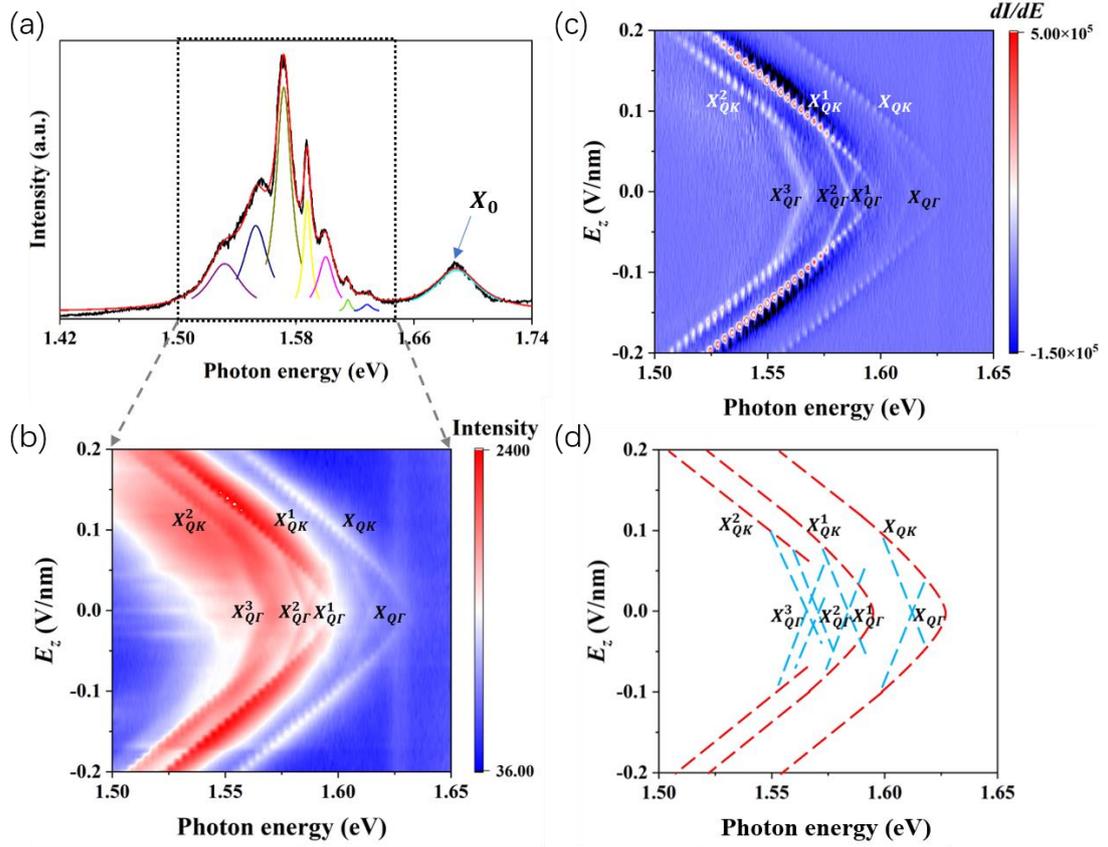

FIG. 2. (a) PL spectrum of device D1 and its fitting under zero gate voltage. (b) Contour plot of the PL spectra as a function of photon energy (bottom axis) and $E_z$ (left axis). $n_0$ remains unchanged. (c) First-order energy derivative of (b). Spatially indirect intervalley excitons are labelled as $X_{QK}$ ($X_{Q\Gamma}$), $X^1_{QK}$ ($X^1_{Q\Gamma}$) and $X^2_{QK}$ ($X^2_{Q\Gamma}$ and $X^3_{Q\Gamma}$) in sequence of decreasing emission energy. (d) Extracted emission energy as a function of $E_z$ from (c) with red ($X_{QK}$ and $X^{1,2}_{QK}$) and blue ($X_{Q\Gamma}$ and $X^{1-3}_{Q\Gamma}$) dashed lines.

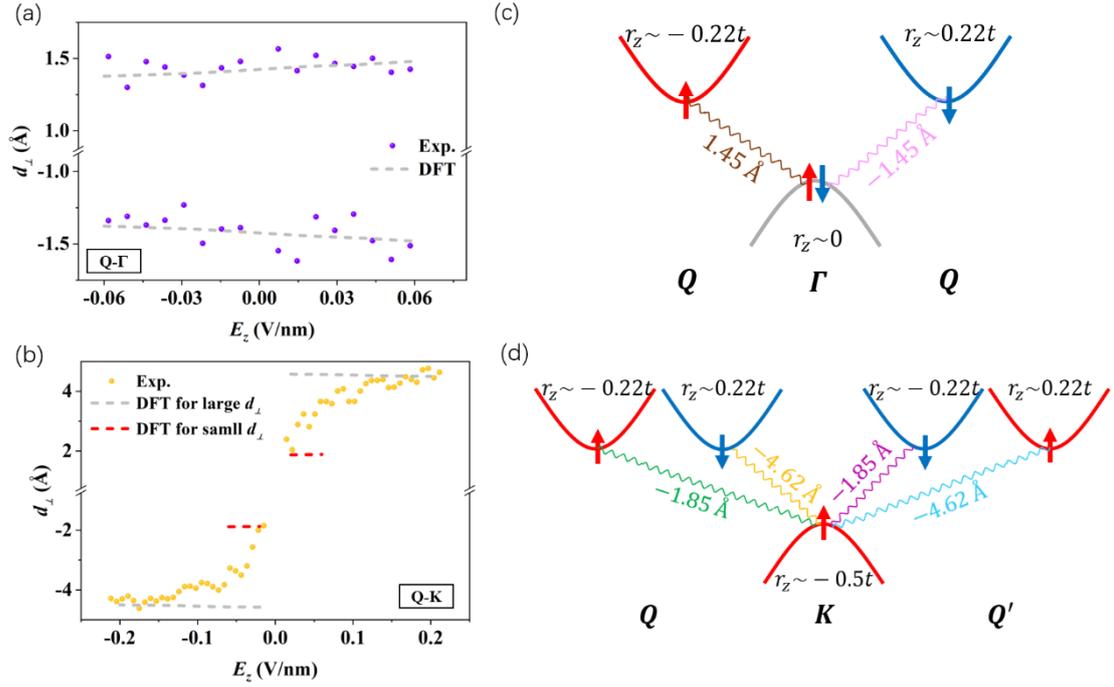

FIG. 3. (a) Experimental (purple dots) and calculated (dashed lines) vertical displacements of $X_{Q\Gamma}$ as a function of $E_z$. (b) Experimental (yellow dots) and theoretically calculated (dashed lines) vertical displacements of $X_{QK}$ versus $E_z$. (c) and (d) Possible transition configurations of $X_{Q\Gamma}$ (c) and $X_{QK}$ (d). $d_\perp$ of each configuration is denoted. Red (blue) curves represent spin-up (spin-down) bands. Valence $\Gamma$ band is spin-degenerated. $r_z$ denotes the equivalent position of wavefunction at zero electric field.

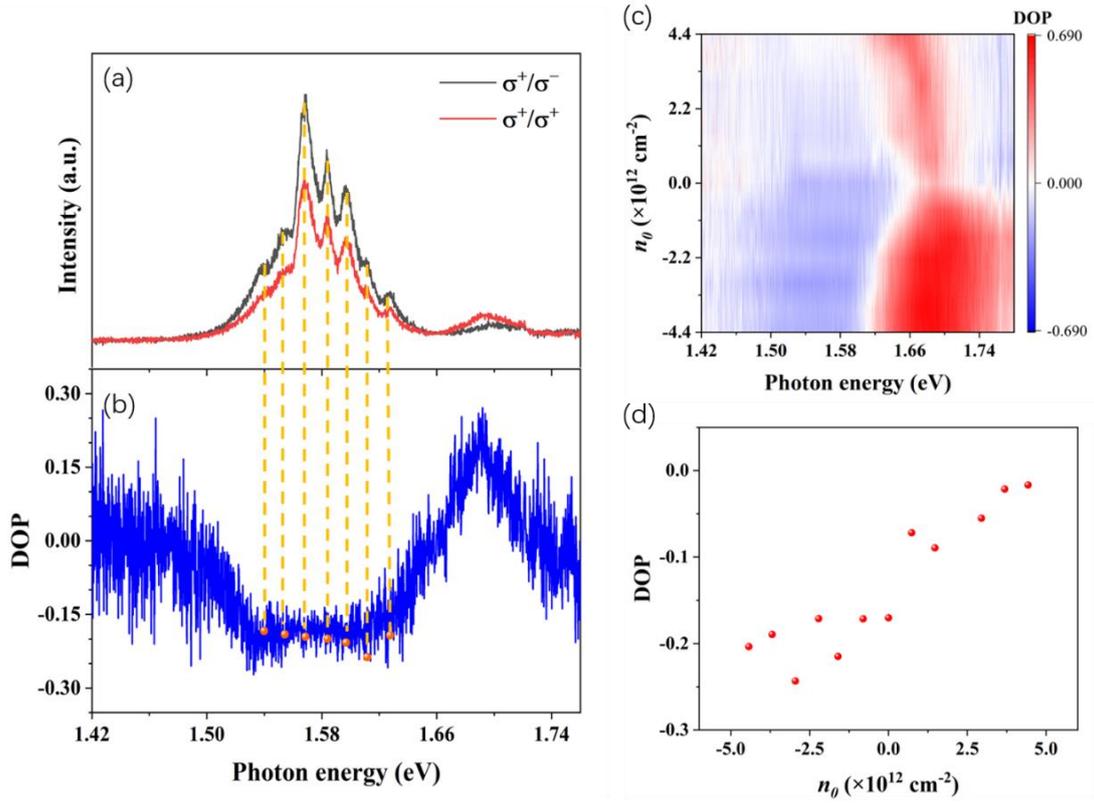

FIG. 4. (a) PL spectra of device D2 under $\sigma^+$ (red line) and $\sigma^-$ (black line) detections, excited by $\sigma^+$ light. (b) The DOP corresponding to (a) as a function of emission energy. DOPs calculated from the measured (blue line) and fitted intensities (orange dots) agree well with each other. (c) Contour plot of the DOP as a function of photon energy (bottom axis) and $n_0$ (left axis). $E_z$ remains unchanged. (d) The DOP versus $n_0$, calculated from the integral intensity in the energy range from 1.45 eV to 1.60 eV. $n_0$ denotes the doping density induced by gate voltage.